\documentstyle[12pt]{article}

\newcommand{\app}[4]{{\em #1} - Acta~Phys.~Polon.       {\bf #2},  #3,  (19#4)}

\newcommand{\am}[4]{{\em #1} - Ann.~Math.               {\bf #2},  #3,  (19#4)}
\newcommand{\ap}[4]{{\em #1} - Ann.~Phys.               {\bf #2},  #3,  (19#4)}

\newcommand{\cmp}[4]{{\em #1} - Commun.~Math.~Phys.     {\bf #2},  #3,  (19#4)}

\newcommand{\ejp}[4]{{\em #1} - Eur.~J.~Phys.           {\bf #2},  #3,  (19#4)}

\newcommand{\ijtp}[4]{{\em #1} - Int.~J.~Theor.~Phys.   {\bf #2},  #3,  (19#4)}

\newcommand{\jmp}[4]{{\em #1} - J.~Math.~Phys.          {\bf #2},  #3,  (19#4)}
\newcommand{\jp}[4]{{\em #1} - J.~Phys.                 {\bf #2},  #3,  (19#4)}

\newcommand{\mpl}[4]{{\em #1} - Mod.~Phys.~Lett.        {\bf #2},  #3,  (19#4)}

\newcommand{\np}[4]{{\em #1} - Nucl.~Phys.              {\bf #2},  #3,  (19#4)}

\newcommand{\nc}[4]{{\em #1} - Nuovo Cim.               {\bf #2},  #3,  (19#4)}

\newcommand{\pess}[4]{{\em #1} - Phys.~Essay            {\bf #2},  #3,  (19#4)}

\newcommand{\pr}[4]{{\em #1} - Phys.~Rev.               {\bf #2},  #3,  (19#4)}
\newcommand{\prl}[4]{{\em #1} - Phys.~Rev.~Lett.        {\bf #2},  #3,  (19#4)}

\newcommand{\ptp}[4]{{\em #1} - Progr.~Theor.~Phys.     {\bf #2},  #3,  (19#4)}

\newcommand{\xx}[3]{                                    {\bf #1},  #2,  (19#3)}

\newcommand{\bel}[1]{\begin{equation}\label{#1}}
\newcommand{\be}{\begin{equation}}
\newcommand{\ee}{\end{equation}}

\newcommand{\beal}[1]{\begin{eqnarray}\label{#1}}
\newcommand{\bea}{\begin{eqnarray}}
\newcommand{\eea}{\end{eqnarray}}

\newcommand{\bean}{\begin{eqnarray*}}
\newcommand{\eean}{\end{eqnarray*}}

\newcommand{\ba}{\begin{array}}
\newcommand{\ea}{\end{array}}

\newcommand{\dv}{\partial}

\newcommand{\bad}[2]{\left( \begin{array}{c}{#1}\\{#2}\end{array} \right)}

\newcommand{\bamq}[4]{\left( \begin{array}{cccc}{#1}&{#2}&{#3}&{#4}\\}
\newcommand{\bamc}[5]{\left( \begin{array}{ccccc}{#1}&{#2}&{#3}&{#4}&{#5}\\}
\newcommand{\eam}{\end{array} \right)}

\newcommand{\ag}{\alpha}
\newcommand{\bg}{\beta}
\newcommand{\cg}{\gamma}
\newcommand{\dg}{\delta}

\newcommand{\og}{\omega}

\newcommand{\Lg}{\Lambda}

\newcommand{\Lb}{\Large \bf}

\newcommand{\fs}{\footnotesize}

\newcommand{\bii}{\begin{itemize}}
\newcommand{\eii}{\end{itemize}}

\newcommand{\ben}{\begin{enumerate}}
\newcommand{\een}{\end{enumerate}}

\newcommand{\bq}{\begin{quote}}
\newcommand{\eq}{\end{quote}}

\newcommand{\bc}{\begin{center}}
\newcommand{\ec}{\end{center}}

\newcommand{\bt}{\begin{tabular}}
\newcommand{\et}{\end{tabular}}

\newcommand{\br}{\begin{flushright}}
\newcommand{\er}{\end{flushright}}

\newcommand{\bl}{\begin{flushleft}}
\newcommand{\el}{\end{flushleft}}

\newcommand{\f}[1]{\footnote{#1}}

\newcommand{\vs}[1]{\vspace*{#1}}

\newcommand{\new}{\pagebreak}

\newcommand{\bb}{\begin{thebibliography}{99}}
\newcommand{\eb}{\end{thebibliography}}
\newcommand{\bi}{\bibitem}

\newcommand{\btp}{\begin{titlepage}}
\newcommand{\etp}{\end{titlepage}}

\newcommand{\go}{\section{Introduction}}
\newcommand{\con}{\section{Conclusions}}
\newcommand{\ack}{\section*{Acknowledgments}}

\begin{document}

\hyphenation{spe-ci-fy qua-ter-ni-ons qua-ter-ni-on qua-ter-nio-nic geo-me-try
super-im-po-se su-per-im-po-si-tion over-co-me opera-tors appro-pria-te
transfor-mation interferen-ces equiva-lent}

\btp

\bc
August, 1995 \hfill {\fs hep-th/95mmnnn}

\vs{3cm}

{\Lb ONE COMPONENT DIRAC EQUATION}

\vs{2cm}

{\sc Stefano De Leo}$^{ \; a)}$

\vs{1cm}

{\it Universit\`a  di Lecce, Dipartimento di Fisica\\
Instituto di Fisica Nucleare, Sezione di Lecce\\
Lecce, 73100, ITALY}

\vs{2cm}

{\bf Abstract}
\ec

\noindent We develop a relativistic free wave equation on the complexified
quaternions, linear in the derivatives. Even if the wave functions are only
one-component, we show that four independent solutions, corresponding to
those of the Dirac equation, exist. A partial set of translations between
complex and complexified quaternionic quantum mechanics may be defined.

\vs{2cm}

\noindent{\fs a) e-mail address:} {\fs \sl deleos@le.infn.it}

\etp

\new

\go

In physics and particularly in quantum mechanics we are
accustomed to distinguishing between ``states'' and ``operators''. Even when
the operators are represented by numerical matrices, the squared form of
operators distinguishes them from the column structure of the spinor states.
Only for one-component fields and operators is there
potential confusion.\\
In extending quantum mechanics defined over the
complex field to quaternions~\cite{fin,adl,adl1} or even complexified
quaternions~\cite{mor} it has almost
always been assumed that matrix operators contain elements which
are ``numbers'', indistinguishable from those of the state vectors. This is
an unjustified limitation.

As the title of this paper indicates, we shall display a one-component Dirac
equation which has only the standard solutions and thus avoids the doubling of
solutions found by other authors who used complexified quaternions. This is
achieved by the introduction of $\cal H$-{\em real linear complexified
quaternions} within operators.

After defining quaternions, complexified quaternions and their
generalizations we will give the simplest example of
their use (namely the one-component Dirac equation) and will introduce
another fundamental ingredient, that is the scalar product, in order to
formulate quantum mechanics.

In 1843, Hamilton~\cite{ham}, after more than a decade of attempts to
generalize complex numbers (seen as a representation of rotations in two
dimensions) in order to describe rotations in three dimensions, discovered
quaternions. Instead of an entity which
he expected to be characterized by three real
numbers, the Irish physicist found that four real numbers were required. He
wrote a quaternion such as\f{In this paper we will use the symbols
$\cal I$, $\cal J$, $\cal K$ instead of the common $i$, $j$, $k$, since we
believe useful to distinguish the {\em quaternionic} imaginary units from
the {\em complex} imaginary unit $i$ which appears in standard quantum
mechanics.}
\[
q=a+{\cal I}b+{\cal J}c+{\cal K}d \; \; \; \; \; , \; \; \; \; \;
a, \; b, \; c, \; d \; \in \; {\cal R} \; \; ,
\]
with operations of multiplication defined according to the following rules
for the imaginary units
\bean
{\cal I}^{2}={\cal J}^{2}={\cal K}^{2} & = & -1\\
{\cal J}{\cal K}=-{\cal K}{\cal J}     & = & {\cal I}\\
{\cal K}{\cal I}=-{\cal I}{\cal K}     & = & {\cal J}\\
{\cal I}{\cal J}=-{\cal J}{\cal I}     & = & {\cal K} \; \; .
\eean
We can immediately extend the previous algebra by introducing a quantity
$i$ whose square is $-1$ and commutes with ${\cal I}, \; {\cal J}, \; {\cal K}$
\bel{xx} q_{\cal C} = \ag +{\cal I}\bg +{\cal J}\cg +{\cal K}\dg  \; \; \;
\; \; ,\; \; \; \; \; \ag , \; \bg , \; \cg , \; \dg \; \in \;
{\cal C}(1, \; i) \; \; . \ee
Complexified quaternions~(\ref{xx}) were introduced in physics to rederive,
more elegantly, all the expressions of special relativity. A review by
Synge~\cite{syn} covers developments up to the 1960s, modern presentations
have been given by Sachs~\cite{sac}, Gough~\cite{gou} and Abonyi~\cite{abo}.
Quantum mechanics on the complexified quaternions is formulated by
Morita~\cite{mor}, in his papers we can find an interesting quaternionic
version of the electroweak theory.\\
Remembering the noncommutativity of the quaternionic multiplication we must
specify whether the quaternionic Hilbert space $V_{{\cal H}_{\cal C}}$ is
to be formed by right or by left multiplication of vectors by scalars.
Besides we must specify if our scalars are quaternionic, complex or real
numbers. We will follow the usual choice (see Morita~\cite{mor}) to work
with a linear vector space under multiplication by complex
(${\cal C}(1, \; i)$) scalars and so the two different conventions,
represented by right or by left multiplication, give isomorphic versions of
the theory.

In the next section we briefly recall the complexified quaternionic algebra
and introduce an appropriate scalar product (or geometry as called by
Rembieli\`nski~\cite{rem}). In the third section we explicitly give a
one-component formulation of the Dirac equation, even if the wave function
is only one-component, we show that four $i$-complex independent solutions
appear. In the following section we identify a set of rules which
permit translations between complex and complexified quaternionic quantum
mechanics. Our conclusion are drawn in the final section.

\section{Complexified quaternionic algebra and $\cal H$-real geometry}

Working with complexified quaternionic numbers we have different
opportunities to define conjugation operations on the imaginary units:
\bc
\bt{|c|c|c|c||c|} \hline
{\fs number}        &       &     &        & \\
{\fs of}            & $\cal I$   & $\cal J$ & $\cal K$    & $i$\\
{\fs conjugations}  &       &     &        & \\ \hline
                    &  $-$  & $+$ & $+$    & \\
    1               &  $+$  & $-$ & $+$    & $+$\\
                    &  $+$  & $+$ & $-$    & \\ \hline
                    &  $+$  & $-$ & $-$    & \\
    2               &  $-$  & $+$ & $-$    & $+$\\
                    &  $-$  & $-$ & $+$    & \\ \hline
    3               &  $-$  & $-$ & $-$    & $+$\\ \hline
\et
\ec
Nevertheless the conjugations expressed in the previous chart are not
independent, in fact we can prove that the conjugation operations which
change only one imaginary unit are connected between themselves and with the
conjugation of all the three imaginary units through similarity
transformations, as in
\[ -{\cal I}\bg +{\cal J}\cg +{\cal K}\dg = -{\cal K}
({\cal I}\bg -{\cal J}\cg +{\cal K}\dg){\cal K} \; \; ,\]
\[ -{\cal I}\bg +{\cal J}\cg +{\cal K}\dg = -{\cal I}
(-{\cal I}\bg -{\cal J}\cg -{\cal K}\dg){\cal I} \; \; .\]
An analogous observation can be formulated for the conjugation of two
imaginary units
\[ -{\cal I}\bg -{\cal J}\cg +{\cal K}\dg = -{\cal I}
(-{\cal I}\bg +{\cal J}\cg -{\cal K}\dg){\cal I} \; \; ,\]
\[ -{\cal I}\bg -{\cal J}\cg +{\cal K}\dg = -{\cal K}
({\cal I}\bg +{\cal J}\cg +{\cal K}\dg){\cal K} \; \; .\]
If we now conjugate the imaginary unit $i$, three possible
conjugations appear
\bc
\bt{|c|c|c|c|c|} \hline
{\fs symbol}            &  $\cal I$  & $\cal J$ & $\cal K$ & $i$    \\ \hline
   $q_{\cal C}^{\; \star}$    &  $-$  & $-$ & $-$ & $+$         \\ \hline
   $q_{\cal C}^{\; *}$        &  $+$  & $+$ & $+$ & $-$         \\ \hline
   $q_{\cal C}^{\; +}$        &  $-$  & $-$ & $-$ & $-$         \\ \hline
\et
\ec
It is straightforward to prove that
\[ (q_{\cal C} p_{\cal C})^{\star}=p_{\cal C}^{\; \star}
q_{\cal C}^{\; \star} \]
and
\[ (q_{\cal C} p_{\cal C})^{*}=q_{\cal C}^{\; *} p_{\cal C}^{\; *} \; \; ,\]
so we immediately have
\[ (q_{\cal C} p_{\cal C})^{+}=(q_{\cal C} p_{\cal C})^{* \star}=
(q_{\cal C}^{\; *} p_{\cal C}^{\; *})^{\star}=
p_{\cal C}^{\; * \star} q_{\cal C}^{\; * \star}=
p_{\cal C}^{\; +} q_{\cal C}^{\; +} \; \; .\]

In quantum mechanics probability amplitudes, rather than probabilities,
superimpose, so we must determine what kinds of number system can be used
for the probability amplitudes $\cal A$. We need a real modulus function
$N({\cal A})$ such that
\[ Probability \; = [N({\cal A})]^{2} \; \; . \]
The first four assumptions on the modulus function are basically technical in
nature
\[ N(0)=0 \; \; , \]
\[ N({\cal A})>0 \; \; if \; \; {\cal A} \neq 0 \; \; , \]
\[ N(r{\cal A})= \vert r \vert N({\cal A}) \; \; \; , \; \; \; r \; \; real
\; \;, \]
\[ N({\cal A}_{1}+{\cal A}_{2}) \leq N({\cal A}_{1}) + N({\cal A}_{2})
\; \; . \]
A final assumption about $N({\cal A})$ is physically motived by imposing the
{\em correspondence principle} in the following form: We require that in
the absence of quantum interferences effects, probability amplitude
superimposition should reduce to probability superimposition. So we have an
additional condition on  $N({\cal A})$:
\[ N({\cal A}_{1} {\cal A}_{2})= N({\cal A}_{1})N({\cal A}_{2}) \; \; . \]
A remarkable theorem of Albert~\cite{alb} shows that the only algebras over
the reals, admitting a modulus functions with the previous properties are
the reals $\cal R$, the complex $\cal C$, the real quaternions
${\cal H}_{\cal R}$ and the octonions $\cal O$.\\
The previous properties
of the modulus function seem to constrain us to work with
{\em division algebras} (which are finite
dimensional algebras for which $a\neq 0$, $b\neq 0$ imply $ab\neq 0$), in
fact
\[ {\cal A}_{1}\neq 0 \; \; \; , \; \; \; {\cal A}_{2}\neq 0 \]
implies
\[ N({\cal A}_{1} {\cal A}_{2})= N({\cal A}_{1})N({\cal A}_{2})\neq 0 \]
which gives
\[ {\cal A}_{1} {\cal A}_{2} \neq 0  \; \; .\]
A simple example of a nondivision algebra is provided by the algebra of
complexified quaternions since
\[ (1+i{\cal I})(1-i{\cal I})=0 \]
guarantees that there are nonzero divisors of zero.\\
So if the probability amplitudes are assumed to be complexified quaternions we
cannot give a satisfactory probability interpretation. Nevertheless we know
that probability amplitudes are connected to inner products,
so we can overcome the above difficulty by defining an appropriate
scalar product. In order
to obtain the right properties of the modulus functions a
possibility is represented by choosing an $\cal H$-{\em real} ( real with
respect to ${\cal I}, \; {\cal J}, \; {\cal K}$) projection
of the complexified quaternionic scalar
product\f{The {\em barred} operators ${\cal O}\mid q$ act on quaternionic
objects $\phi$ as follows
\[ ({\cal O}\mid q) \phi \: \equiv \: {\cal O} \phi q \; \: .\]}:
\bel{pp}
{\cal A} = \; <\psi \mid \phi>_{{\cal C}(1, \; i)} \; =
\frac{1-{\cal I}\mid {\cal I}-{\cal J} \mid {\cal J} -
{\cal K} \mid {\cal K}}{4} <\psi \mid \phi> \; \; .
\ee
The probability amplitude $\cal A$ (now $i$-complex valued) satisfies the
required properties. In this paper we will treat a complexified
quaternionic quantum mechanics with $\cal H$-real ($i$-complex) geometry.

In the expression~(\ref{pp}) we recognize a particular $\cal H$-{\em real
linear complexified quaternion}. These objects represent the most general
transformation on
complexified quaternions, explicitly such transformations are expressed by
\bel{rlcq}
{\cal O} = q_{\cal C} + p_{\cal C} \mid {\cal I} + r_{\cal C} \mid {\cal J} +
s_{\cal C} \mid {\cal K} \; \; ,
\ee
\[  q_{\cal C} , \; p_{\cal C}, \; r_{\cal C}, \; s_{\cal C}
\in {\cal H}_{\cal C} \; \; .\]
An operator like~(\ref{rlcq}) satisfies the following relation
\[ {\cal O} (\psi \ag) = ({\cal O} \psi) \ag \; \; ,\]
for an arbitrary $\ag \in {\cal C}(1, \; i)$, from here the name of
``$\cal H$-real'' (real with respect to $\cal I$, $\cal J$ and $\cal K$)
or ``$i$-complex'' linear operator.\\
Thanks to  $\cal H$-{\em real linear complexified quaternions}
we will be able to
write a one-component Dirac equation (characterized by four orthogonal
solutions) and will overcome previous problems.

\section{One-component Dirac Equation}

Various formulations of the Dirac relativistic equation on the complexified
quaternionic field have been considered since the 1930s.
A pioneer, in this
field, was certainly Conway~\cite{con}, more recent presentations
can be found in the papers of Edmonds~\cite{edm} and
Gough~\cite{gou,gou2}. When written in this manner a doubling of solutions
from four to eight, occurs.

We briefly recall the ``standard'' complexified quaternionic Dirac equation,
as formulated by Edmonds and Gough. If we represent the energy-momentum by
the following complexified quaternion
\be
{\cal P}= E+i({\cal I}p_{x}+{\cal J}p_{y}+{\cal K}p_{z}) \; \; ,
\ee
a relativistic free wave equation is quickly obtained by the substitution
\be
{\cal P} \; \rightarrow  \; i{\cal D} =
i\partial_{t} + {\cal I}\partial_{x} + {\cal J}\partial_{y} +
{\cal K} \partial_{z}
\ee
in
\bel{cop}
{\cal P}\psi_{a}=m\psi_{b} \; \; \; , \; \; \;
{\cal P}^{\star}\psi_{b}=m\psi_{a} \; \; .
\ee
The spinors $\psi_{a}$ and $\psi_{b}$ satisfy the Klein-Gordon equation, in
fact we have
\bc
\bt{lclcl}
${\cal P}^{\star}{\cal P}\psi_{a}$ & $=$ & $m {\cal P}^{\star}\psi_{b}$ &
$=$ &  $m^{2}\psi_{a} \; \; ,$\\
${\cal P}{\cal P}^{\star}\psi_{b}$ & $=$ &  $m {\cal P}\psi_{a}$ &
$=$ & $m^{2}\psi_{b} \; \; ,$
\et
\ec
\bc
${\cal P}{\cal P}^{\star}={\cal P}^{\star}{\cal P}=-{\partial_{t}}^{2}
+ {\vec{\partial}}^{\; 2} \; \; .$
\ec
The covariance is obtained if, under Lorentz transformations
\[ {\cal P}'=\Lambda {\cal P} \Lambda^{+} \; \; \; \; , \; \; \; \;
\Lambda^{\star} \Lambda = 1 \; \; ,\]
\[ \Lg \in {\cal H}_{\cal C} \; \; ,\]
the spinors $\psi_{a}$ and $\psi_{b}$ transform as follows
\be
{\psi_{a}}'=\Lambda^{*}\psi_{a} \; \; \; , \; \; \;
{\psi_{b}}'=\Lambda \psi_{b} \; \; .
\ee
The equations~(\ref{cop}) can be rewritten in matrix form as follows
\be
\left[ \left( \begin{array}{cc} 0 & 1\\
1 & 0 \end{array} \right) i\partial_{t} +
\left( \begin{array}{cc} 0 & -1\\
1 & 0 \end{array} \right) \vec{Q}\cdot \vec{\partial} \right]
\left( \begin{array}{c} \psi_{a}\\
\psi_{b} \end{array} \right) =
m \left( \begin{array}{c} \psi_{a}\\
\psi_{b} \end{array} \right) \; \; ,
\ee
\[ \vec{Q} \equiv ({\cal I}, \; {\cal J}, \; {\cal K}) \; \; . \]
We recognize, in the complexified quaternionic Dirac equation,
eight orthogonal solutions, whereas the standard Dirac
equation involves only four. For $\vec{p}=0$, we find
{\fs
\bc
\bt{lclclclclc}
$E=+m$ & : & $\frac{1}{\sqrt{2}} \bad{1}{1}$ & , &
$ \frac{1}{\sqrt{2}} \bad{1}{1} \; {\cal I}$ & , &
$ \frac{1}{\sqrt{2}} \bad{1}{1} \; {\cal J}$ & , &
$ \frac{1}{\sqrt{2}} \bad{1}{1} \; {\cal K}$ & ; \\
 & & & & & & & & & \\
$E=-m$ & : & $\frac{1}{\sqrt{2}} \bad{1}{-1}$ & , &
$ \frac{1}{\sqrt{2}} \bad{1}{-1} \; {\cal I}$ & , &
$ \frac{1}{\sqrt{2}} \bad{1}{-1} \; {\cal J}$ & , &
$ \frac{1}{\sqrt{2}} \bad{1}{-1} \; {\cal K}$ & .
\et
\ec
}
Remember that with $\cal H$-real geometry {\fs $\mid \psi >$,
$\mid \psi >{\cal I}$, $\mid \psi >{\cal J}$, $\mid \psi >{\cal K}$}
represent orthogonal states. The possible physical significance of
these additional solutions has been a matter of speculation~\cite{edm2}.

We will show (see section 4) that this doubling of solutions is strictly
connected with the use of reducible matrices
\bel{mo} \left( \begin{array}{cc} 0 & i\\
i & 0 \end{array} \right)  \; \; \; ,  \; \; \;
\left( \begin{array}{cc} 0 & -\vec{Q}\\
\vec{Q} & 0 \end{array} \right) \ee
and so there is no ``new physics'' in the complexified quaternionic Dirac
equation. Finally,
we remark that when one works with complexified quaternions, matrix
operators like~(\ref{mo}) represent an unjustified limitation.

We now derive, following the standard Dirac approach, a {\em one-component}
equation with only {\em four} solutions.\\
In order to obtain a
positive-definite probability density $\rho$, we require an equation linear
in $\partial_{t}$, then, for relativistic covariance, the equation must
also be linear in $\vec{\partial}$. The simplest equation
that we can write down is
\bel{c}
i\dv_{t} \psi = \vec{Q} \cdot \vec{\cal P} \psi {\cal I} + m i \psi
{\cal J} \; \; ,
\ee
with
\[ \vec{\cal P} \equiv - i \vec{\dv} \; \; .\]
Our Hamiltonian, given by
\[ \vec{Q} \cdot \vec{\cal P} \mid {\cal I} + m i \mid {\cal J}
\; \; ,\]
represent a particular $\cal H$-{\em real linear complexified quaternion}.
It is straightforward to prove that $\psi$ satisfies the Klein-Gordon equation,
in fact the operators
\bel{ww}
\vec{\ag}\equiv \vec{Q}\mid {\cal I} \; \; \; , \; \; \;
\bg \equiv i \mid {\cal J} \ee
verify the usual relations\f{Note that
\[ (q_{\cal C} \; \vert \; q)(p_{\cal C} \; \vert \; p)=
q_{\cal C}p_{\cal C} \; \vert \; pq \; \; ,\]
\[ q_{\cal C}, \; p_{\cal C} \; \in \; {\cal H}_{\cal C} \; \; , \; \;
q, \; p \; \in \; {\cal H}_{\cal R} \; \; .\]}
\bean
    \{\vec{\ag}, \; \bg\} =\{\ag_{x}, \; \ag_{y}\}=\{\ag_{x}, \; \ag_{z}\}=
\{\ag_{y}, \; \ag_{z}\}              & = & 0 \; \; ,\\
\ag_{x}^{2}=\ag_{y}^{2}=\ag_{z}^{2}=\bg^{2}                & = & 1 \; \; .
\eean

In order to find a positive-definite probability density, we must consider an
appropriate conjugation, noting that
\bean
\psi^{\star}\psi & = & \psi_{1}^{2} + \psi_{2}^{2} + \psi_{3}^{2}
+\psi_{4}^{2}\\
\psi^{*}\psi & = & \vert \psi_{1} \vert^{2} - \vert \psi_{2} \vert^{2} -
\vert \psi_{3} \vert^{2}  - \vert \psi_{4} \vert^{2} +
({\cal I}, \; {\cal J}, \; {\cal K}) \; \; terms\\
\psi^{+}\psi & = & \vert \psi_{1} \vert^{2} + \vert \psi_{2} \vert^{2} +
\vert \psi_{3} \vert^{2}  + \vert \psi_{4} \vert^{2} +
({\cal I}, \; {\cal J}, \; {\cal K}) \; \; terms
\eean
\[ ( \; \psi=\psi_{1} +{\cal I}\psi_{2}+{\cal J}\psi_{3}+{\cal K}\psi_{4}
\; \; \; ;
\; \; \; \psi_{1}, \; \psi_{2}, \; \psi_{3}, \; \psi_{4} \; \in
\;  {\cal C}(1, \; i) \; ) \; \; ,\]
we choose
\be
\rho = (\psi^{+} \psi)_{{\cal C}(1, \; i)} \; \; .
\ee
We need an $\cal H$-real ($i$-complex) geometry and a $^{+}$ conjugation.
Using such geometry, given an operator like~(\ref{rlcq}), we can immediately
write its hermitian conjugate as
\be
{\cal O}^{+} = q_{\cal C}^{+} - p_{\cal C}^{+} \mid {\cal I} -
r_{\cal C}^{+} \mid {\cal J} - s_{\cal C}^{+} \mid {\cal K} \; \; .
\ee

We now consider a new equivalent representation for the operators
$\vec{\ag}$ and $\bg$, which allows us to simplify the following steps
\bel{b}
\vec{\ag}\equiv (-{\cal J}\mid {\cal I}, \; i\mid {\cal K}, \;
{\cal K} \mid {\cal I}) \; \; \; ,
\; \; \; \bg \equiv -{\cal I}\mid {\cal I} \; \; .
\ee
We seek plane wave solutions of the quaternionic Dirac
equation~(\ref{c}), i.e., solutions of the form
\bc
\bt{ll}
$\psi^{+}(x)=u(p) e^{-i(Et-\vec{p}\cdot \vec{x})}$ &
{\fs positive energy ,}\\
$\psi^{-}(x)=v(p) e^{+i(Et+\vec{p}\cdot \vec{x})}$ &
{\fs negative energy ,}
\et
\ec
with the condition that $E$ is positive.\\
In the rest frame of the particle, $\vec{p}=0$, the Dirac equation reduces
to
\bean
u(m)= & -{\cal I}u(m){\cal I} \; \; ,\\
v(m)= & +{\cal I}v(m){\cal I} \; \; .
\eean
There are two linearly independent $u$ solutions, and two $v$'s, we denote
them as follows
\[ u^{(1)}(m)=1 \; \; , \; \; u^{(2)}(m)={\cal I} \; \; ,
\; \; v^{(1)}(m)={\cal J} \; \;
, \; \; u^{(1)}(m)={\cal K} \; \; .\]
The Dirac equation implies on the spinors $u(p)$ and $v(p)$ the conditions
\bean
(E-\vec{\ag}\cdot \vec{p} - \bg m)u(p) & = & 0 \; \; ,\\
(E+\vec{\ag}\cdot \vec{p} + \bg m)v(p) & = & 0 \; \; ,
\eean
so we may write, for $\vec{p}\neq 0$, the following solutions
\bean
u^{(1, \; 2)}(p) = & (E+\vec{\ag}\cdot \vec{p} + \bg m)u^{(1, \; 2)}(m)
\; \; ,\\
v^{(1, \; 2)}(p) = & (E-\vec{\ag}\cdot \vec{p} - \bg m)v^{(1, \; 2)}(m)
\; \; .
\eean
Explicitly,
\bean
u^{(1)}(p) = & (2m(m+E))^{-\frac{1}{2}}
(E+m+{\cal K}(p_{x}+ip_{y})+{\cal J}p_{z}) \; \; \; \; ,\\
u^{(2)}(p) = & (2m(m+E))^{-\frac{1}{2}}
(E+m+{\cal K}(p_{x}-ip_{y})+{\cal J}p_{z}){\cal I} \; \; ,\\
v^{(1)}(p) = & (2m(m+E))^{-\frac{1}{2}}
(E+m+{\cal K}(p_{x}+ip_{y})+{\cal J}p_{z}){\cal J} \; \; ,\\
v^{(2)}(p) = & (2m(m+E))^{-\frac{1}{2}}
(E+m+{\cal K}(p_{x}-ip_{y})+{\cal J}p_{z}){\cal K}  \; \; .
\eean
The normalization factors have been chosen in order that
\[ u^{(1) \; +}u^{(1)}=u^{(2) \; +}u^{(2)}=
v^{(1) \; +}v^{(1)}=v^{(2) \; +}v^{(2)}=\frac{E}{m} \; \; .\]
The orthogonality of the Dirac solutions is immediately proved
by the following considerations
\[ u^{(2) \; +}=-{\cal I}u^{(1) \; \star} \; \; \; , \; \; \;
v^{(1)}=u^{(1)}{\cal J} \; \; \; , \; \; \; v^{(2)}=u^{(2)}{\cal J}\]
and
\[  u^{(1) \; \star}u^{(1)} \in {\cal C}(1, \; i) \; \; .\]
All non-diagonal products are $\cal H$-imaginary ($\sim {\cal I}, \;
{\cal J}, \;  {\cal K}$) and so our four solutions are orthogonal.\\
If we rewrite the complexified quaternionic Dirac equation in covariant
form
\bel{dcf} i\gamma^{\mu}\partial_{\mu}\psi=m\psi \; \; , \ee
where
\[ \gamma^{\mu} \; \equiv \; (\bg, \; \bg \vec{\ag}) \; \; , \]
we can quickly extract the standard results, since our
$\gamma^{\mu}$-matrices, now $\cal H$-real linear complexified
quaternionic numbers, satisfy the $i$-complex Dirac algebra. For
example, in order to obtain the relativistic covariance we must assume
that, under Lorentz transformations ($x'=\Lg x$), there be a linear
relation between the wave function $\psi$ in the first frame and the wave
function $\psi'$ in the transformed frame, namely
\[ {\psi}' = {\cal T}(\Lg)\psi \; \; ,\]
with
\[ {\cal T}(\Lg) = e^{\frac{1}{8}[\cg^{\mu}, \; \cg^{\nu}]\og_{\mu \nu}} \;
\; ,\]
\bc
{\fs $\og_{\mu \nu}$ antisymmetric in $\mu \nu$}.
\ec
Considering an infinitesimal rotation around $z$ and finding the
corresponding transformation of the wave function $\psi$, we obtain the
spin operator
\be
{\cal S}_{z} = -\frac{1}{2}\;  {\cal J}\mid {\cal J} \; \; ,
\ee
and so our four solutions $u^{(1, \; 2)}, \; v^{(1, \; 2)}$
correspond to positive and negative energy solutions with
${\cal S}=\frac{1}{2}$ and for $\vec{p}=(0, \; 0, \; p_{z})$, to
${\cal S}_{z}=\frac{1}{2}, \; -\frac{1}{2}, \; \frac{1}{2}, \;
-\frac{1}{2}$, respectively.

We can introduce in the Dirac equation a potential $A^{\mu}$ by the minimal
coupling:
\[ i\partial^{\mu} \rightarrow i\partial^{\mu} -
eA^{\mu} \; \; ,\]
\bc
{\fs where $e=-\vert e \vert$  for the electron}.
\ec
We will seek a transformation $\psi \rightarrow \psi_{c}$ reversing the
charge, such that
\bel{p}
(i\partial^{\mu} -eA^{\mu})\gamma_{\mu}\psi=m\psi \; \; ,
\ee
\be
(i\partial^{\mu} +eA^{\mu})\gamma_{\mu}\psi_{c}=m\psi_{c} \; \; .
\ee
To construct $\psi_{c}$ we conjugate (* operation) the first equation and
multiply it for $i \; \vert \; j$
\[ ({\cal I} \; \vert \; {\cal J})(-i\partial^{\mu} -eA^{\mu})
\gamma_{\mu}^{*}\psi^{*} = m {\cal I} \psi^{*} {\cal J} \; \; .\]
In our representation, it is straightforward to prove that
\[ ({\cal I} \; \vert \; {\cal J})\gamma_{\mu}^{*}\psi^{*} =
- \gamma_{\mu}{\cal I} \psi^{*} {\cal J} \]
and thus
\be
\psi_{c} = {\cal I} \psi^{*} {\cal J} \; \; ,
\ee
which represents the quaternionic version of the charge conjugation
operation.

We conclude this section with the following consideration:\\
Comparing the equation~(\ref{dcf}) with the standard Dirac equation
\[ i\gamma^{\mu}\partial_{\mu}\psi=m\psi \; \; , \]
where
\[ \gamma^{0} = \left( \begin{array}{cc} 1 & \cdot\\
\cdot & -1 \end{array} \right) \; \; \; , \; \; \;
\vec{\gamma} = \left( \begin{array}{cc} \cdot & \vec{\sigma}\\
-\vec{\sigma} & \cdot \end{array} \right) \; \; , \]
we can obtain an interesting relation between $4\times 4$ complex
matrices and $\cal H$-real linear complexified quaternions
\bel{id}
i\gamma^{0} \; \; \leftrightarrow \; \; -i{\cal I} \; \vert \; {\cal I}\; \;
\; \; , \; \; \; \; i\vec{\gamma} ; \; \leftrightarrow \; \;
(-i{\cal K}, \; {\cal I} \; \vert \; {\cal J}, \; -i{\cal J}) \; \; .
\ee
With this as encouragement we will derive a complete translation in the
following section.

\section{A possible translation}

In this section we give explicitly a set of rules for passing back and
forth between standard (complex) quantum mechanics and our complexified
quaternionic version. Nevertheless we must remark that this translation will
not be possible in all situations, so it is only partial. We will be able to
pass from $4n$ dimensional complex matrices to $n$ dimensional $\cal H$-real
linear complexified quaternionic matrices.

We know that sixteen complex numbers are necessary to define
the most general $4\times 4$ complex matrix but only four are needed to
define the most general complexified quaternion
\[ \psi=\psi_{1} +{\cal I}\psi_{2}+{\cal J}\psi_{3}+{\cal K}\psi_{4} \]
\[ ( \; \psi_{1}, \; \psi_{2}, \; \psi_{3}, \; \psi_{4} \; \in
\;  {\cal C}(1, \; i) \; ) \; \; .\]
Therefore, in order to achieve a translation, we need twelve new complex
${\cal C}(1, \; i)$ numbers. If we work with $\cal H$-real linear complexified
quaternions, these new degrees of freedom are represented by
\[ \phi \mid {\cal I} + \eta \mid {\cal J} + \xi \mid {\cal K} \]
\[ ( \phi, \; \eta, \; \xi \; \in \; {\cal H}_{\cal C}) \; \; . \]

The rules to translate between complex and complexified quaternionic
quantum mechanics are:
\[ i \leftrightarrow \left( \begin{array}{cccc} i & 0 & 0 & 0\\
0 & i & 0 & 0\\ 0 & 0 & i & 0\\ 0 & 0 & 0 & i\end{array} \right) \; \; , \]
\bean
{\cal I} \leftrightarrow \left( \begin{array}{cccc} 0 & $-1$ & 0 & 0\\
1 & 0 & 0 & 0\\ 0 & 0 & 0 & $-1$\\ 0 & 0 & 1 & 0\end{array} \right) & , &
1 \; \vert \; {\cal I} \leftrightarrow
\left( \begin{array}{cccc} 0 & $-1$ & 0 & 0\\
1 & 0 & 0 & 0\\ 0 & 0 & 0 & 1\\ 0 & 0 & $-1$ & 0\end{array} \right) \; \; ,\\
\\
{\cal J} \leftrightarrow \left( \begin{array}{cccc} 0 & 0 & $-1$ & 0\\
0 & 0 & 0 & 1\\ 1 & 0 & 0 & 0\\ 0 & $-1$ & 0 & 0\end{array} \right) & , &
1 \; \vert \; {\cal J} \leftrightarrow
\left( \begin{array}{cccc} 0 & 0 & $-1$ & 0\\
0 & 0 & 0 & $-1$\\ 1 & 0 & 0 & 0\\ 0 & 1 & 0 & 0\end{array} \right) \; \; .
\eean
We can obviously obtain the remaining rules from the previous ones, for example
\bean
{\cal K}={\cal I}{\cal J}  & \leftrightarrow  &
\left( \begin{array}{cccc} 0 & 0 & 0 & $-1$\\
0 & 0 & $-1$ & 0\\ 0 & 1 & 0 & 0\\ 1 & 0 & 0 & 0\end{array}
\right) \; \; , \\ \\
1\mid {\cal K}= (1 \mid {\cal J}) (1 \mid {\cal I}) & \leftrightarrow &
\left( \begin{array}{cccc} 0 & 0 & 0 & $-1$\\
0 & 0 & 1 & 0\\ 0 & $-1$ & 0 & 0\\ 1 & 0 & 0 & 0\end{array} \right) \; \; .
\eean
With these rules we can translate any $4n$ dimensional complex matrix into
an equivalent $n$ dimensional $\cal H$-real linear complexified matrix and
{\em viceversa}
\be
\left( \begin{array}{cccc} a_{1} & b_{1} & c_{1} & d_{1}\\
a_{2} & b_{2} & c_{2} & d_{2}\\ a_{3} & b_{3} & c_{3} & d_{3}\\
a_{4} & b_{4} & c_{4} & d_{4}\end{array} \right) \;
\leftrightarrow \; q_{\cal C} +p_{\cal C} \; \vert \; {\cal I} +
r_{\cal C} \; \vert \; {\cal J} + s_{\cal C} \; \vert  \; {\cal K}
\; \; ,
\ee
with
\bean
4q_{\cal C} & = & a_{1}+b_{2}+c_{3}+d_{4}+{\cal I}(a_{2}-b_{1}+c_{4}-d_{3})+\\
   &   & +{\cal J}(a_{3}-b_{4}-c_{1}+d_{2})+{\cal K}(a_{4}+b_{3}-c_{2}-d_{1})
\; \; ,\\
4p_{\cal C} & = & a_{2}-b_{1}-c_{4}+d_{3}-{\cal I}(a_{1}+b_{2}-c_{3}-d_{4})+\\
   &   & -{\cal J}(a_{4}+b_{3}+c_{2}+d_{1})+{\cal K}(a_{3}-b_{4}+c_{1}-d_{2})
\; \; ,\\
4r_{\cal C} & = & a_{3}+b_{4}-c_{1}-d_{2}+{\cal I}(a_{4}-b_{3}-c_{2}+d_{1})+\\
   &   & -{\cal J}(a_{1}-b_{2}+c_{3}-d_{4})-{\cal K}(a_{2}+b_{1}+c_{4}+d_{3})
\; \; ,\\
4s_{\cal C} & = & a_{4}-b_{3}+c_{2}-d_{1}-{\cal I}(a_{3}+b_{4}+c_{1}+d_{2})+\\
   &   & +{\cal J}(a_{2}+b_{1}-c_{4}-d_{3})-{\cal K}(a_{1}-b_{2}-c_{3}+d_{4})
\; \; .
\eean
We can obviously invert the previous development and obtain, given
$q_{\cal C}, \; p_{\cal C}, \; r_{\cal C}, \; s_{\cal C}$ the complex
coefficients $a, \; b, \; c, \; d$, i.e., as in
\bean
a_{1} & = & \left(
q_{\cal C}+{\cal I}p_{\cal C}+{\cal J}r_{\cal C}+{\cal K}s_{\cal C}
\right)_{{\cal C}(1, \; i)} \; \; ,\\
c_{2} & = & \left(
s_{\cal C}+{\cal I}r_{\cal C}+{\cal J}p_{\cal C}+{\cal K}q_{\cal C}
\right)_{{\cal C}(1, \; i)} \; \; .
\eean
We now give an example of this translation. Consider the spin operator
which appear in the standard Dirac equation (see for example~\cite{itz}):
\[S_{z} = \frac{1}{2} \left( \begin{array}{cccc} 1 & 0 & 0 & 0\\
0 & $-1$ & 0 & 0\\ 0 & 0 & 1 & 0\\ 0 & 0 & 0 & $-1$\end{array} \right) \; \;
.\]
The only non zero coefficients are $a_{1}=c_{3}=-b_{2}=-d_{4}=1$ and so
the complexified quaternionic  version of the spin operator is
\[S_{z} =- \frac{1}{2} \; {\cal J} \; \vert \; {\cal J} \; \; .\]
In similar way we can obtain all the results given in the third section, we
must only translate from standard (complex) Dirac equation, so we cannot
have a doubling of the solutions.

We conclude this section by giving the matrix $\cal S$ which
reduces the complexified quaternionic matrices used by Edmonds~\cite{edm} and
Gough~\cite{gou}, in this matter we have elegantly
explained the doubling of
solutions in their papers\footnote{In the recent article of
ref.~\cite{and} we read the following comment about the complexified
quaternionic
version of Dirac equation: ``{\sl When written in this manner a doubling of
components of the wave function from $4$ to $8$ occurs.}'' This difficulty
is now overcome.}.\\
The matrix ${\cal S}$ which satisfies
\[{\cal S} \; i\left( \begin{array}{cc} 0 & 1\\
1 & 0\end{array} \right) \; {\cal S}^{-1} = (1 \; \vert \; {\cal J}) \;
\left( \begin{array}{cc} 1 & 0\\
0 & 1\end{array} \right) \; \; ,\]
\[{\cal S} \; \vec{Q}\left( \begin{array}{cc} 0 & $-1$\\
1 & 0\end{array} \right) \; {\cal S}^{-1} = (\vec{Q} \; \vert \; {\cal I})
\left( \begin{array}{cc} 1 & 0\\
0 & 1\end{array} \right) \; \; ,\]
is
\[{\cal S} \; = \; \frac{1}{2\sqrt{2}} \; \left( \begin{array}{cc}
1+i \; \vert \; {\cal I} -1 \; \vert \; {\cal J} -i \; \vert \; {\cal K}
\; \; \; &
\; \; \; -i +1 \; \vert \; {\cal I} -i \; \vert \; {\cal J} +
1 \; \vert \; {\cal K}\\
 & \\
i+1 \; \vert \; {\cal I} +i \; \vert \; {\cal J} +1 \; \vert \; {\cal K}
\; \; \; &
\; \; \;-1 +i \; \vert \; {\cal I} +1 \; \vert \; {\cal J} -
i \; \vert \; {\cal K}
\end{array} \right) \; \; .\]

\con

In this paper we have defined a set of rules for translating from standard
(complex) quantum mechanics to complexified quaternionic quantum mechanics,
with our rules we can obtain a rapid counterpart of the standard quantum
mechanical results and overcome previous difficulties (like doubling of the
solutions in the Dirac equation).\\
We remark that there is no ``new physics'' in the complexified
quaternionic Dirac equation, in contrast with the standard folklore.
Nevertheless we emphasize that our translation is only a partial translation,
so observable differences could be found between complex and complexified
quaternionic quantum mechanics.

\ack

The author thanks Pietro Rotelli for useful comments and suggestions.

\bb
\bi{fin}
\jmp{Finkelstein D, Jauch JM, et al.}{3}{207}{62}; \xx{4}{788}{63}.\\
\jmp{Finkelstein D, Jauch JM, Speiser D}{4}{136}{63}.\\
{\em Finkelstein D, Jauch JM, Speiser D} - {\sl Notes on quaternion quantum
mechanics}\\
in Logico-Algebraic Approach to Quantum Mechanics II, Hooker (Reidel,
Dordrecht 1979), 367-421.
\bi{adl}
\pr{Adler SL}{D21}{550}{80}, \xx{D21}{2903}{80};\\
\prl{Adler SL}{55}{783}{85}; Phys.~Rev. \xx{D34}{1871}{86};\\
\cmp{Adler SL}{104}{611}{86};\\
\pr{Adler SL}{D37}{3654}{88}; Phys.~Lett. \xx{B221}{39}{89};\\
\np{Adler SL}{B415}{195}{94}.\\
\pr{Davies AJ}{D41}{2628}{90}.\\
\pr{De Leo S, Rotelli P}{D45}{575}{92};\\
\ptp{De Leo S, Rotelli P}{92}{917}{94};\\
\nc{De Leo S , Rotelli P}{B110}{33}{95};\\
{\em De Leo S, Rotelli P} - {\sl Quaternion Higgs and the Electroweak Gauge
Group}\\ {\fs to appear in {\bf Int.~J.~Mod.~Phys.~A}}.\\
\ap{Horwitz LP,  Biedenharn LC}{157}{432}{84}.\\
\app{Horwitz LP, Razon A}{24}{141}{91}; \xx{24}{179}{91};\\
\jmp{Horwitz LP, Razon A}{33}{3098}{92}.\\
\jmp{Horwitz LP}{34}{3405}{93}.\\
\jmp{Nash CC, Joshi GC}{28}{2883}{87}; \xx{28}{2886}{87};\\
\ijtp{Nash CC, Joshi GC}{27}{409}{88}; \xx{31}{965}{92}.\\
\mpl{Rotelli P}{A4}{933}{89}; \xx{A4}{1763}{89}.
\bi{adl1}
{\em Adler SL} - {\sl Quaternion quantum mechanics and quantum fields}\\
Oxford UP, Oxford, 1995.
\bi{mor}
\ptp{Morita K}{67}{1860}{81}, \xx{68}{2159}{82};\\
\ptp{Morita K}{70}{1648}{83}; \xx{72}{1056}{84};\\
\ptp{Morita K}{73}{999}{84}; \xx{75}{220}{85};\\
\ptp{Morita K}{90}{219}{93}.
\bi{ham}
{\em Hamilton WR} - {\sl Elements of Quaternions} - Chelsea Publishing,
N.~Y.~, 1969.
\bi{syn}
{\em Synge JL} - {\sl Quaternions, Lorentz Transformation, and the Conway Dirac
Eddington Matrices} - Institute for Advanced Study, Dublin, 1972.
\bi{sac}
{\em Sachs M} - {\sl General Relativity and Matter: A spinor Field Theory
from Fermis to Light-Years} - D.~Reidel, Dordrecht, 1982.
\bi{gou}
\ejp{Gough W}{10}{188}{89}.
\bi{abo}
\jp{Abonyi I, Bito JF, Tar JK}{A24}{3245}{91}.
\bi{rem}
\jp{Rembieli\`nski J}{A11}{2323}{78}.
\bi{alb}
\am{Albert AA}{48}{495}{47}.
\bi{con}
{\em Conway AW} - Pro.~R.~Soc. {\bf A162}, 145 (1937).
\bi{edm}
\ijtp{Edmonds JD}{6}{205}{72}; Am.~J.~Phys. \xx{42}{220}{74}.
\bi{gou2}
\ejp{Gough W}{7}{35}{86}; \xx{8}{164}{87}.
\bi{edm2}
{\em Edmonds JD} - Found.~of Phys. {\bf 3}, 313 (1973).
\bi{itz}
{\em Itzykson C, Zuber JB} - {\sl Quantum Field Theory} - McGraw-Hill Book
Co, Singapore, 1985.
\bi{and}
\pess{Anderson R, Joshi GC}{6}{308}{93}.
\eb

\end{document}